\documentclass[longbibliography,twocolumn,prl,aps,superscriptaddress,showpacs,amsmath,amssymb,floatfix]{revtex4-1}
\usepackage{color}
\usepackage{xcolor}
\usepackage{mathrsfs}
\usepackage{amsmath}
\usepackage{graphicx}
\usepackage{dcolumn}
\usepackage{bm}
\usepackage{amssymb}
\usepackage{float}
\usepackage{array}
\usepackage{tikz}
\usepackage{ulem}

\begin{document}

\title{Emergent Haldane phase in an alternating bond $\mathbb{Z}_3$ parafermion chain}
\author{Shun-Yao Zhang}
\affiliation{CAS Key Laboratory of Quantum Information, University of Science and Technology of China, Hefei, 230026, People’s Republic of China}
\author{Hong-Ze Xu}
\affiliation{CAS Key Laboratory of Quantum Information, University of Science and Technology of China, Hefei, 230026, People’s Republic of China}
\author{Yue-Xin Huang}
\affiliation{CAS Key Laboratory of Quantum Information, University of Science and Technology of China, Hefei, 230026, People’s Republic of China}
\author{Guang-Can Guo}
\affiliation{CAS Key Laboratory of Quantum Information, University of Science and Technology of China, Hefei, 230026, People’s Republic of China}
\affiliation {Synergetic Innovation Center of Quantum Information and Quantum Physics, University of Science and Technology of China, Hefei, Anhui 230026, China}
\author{Zheng-Wei Zhou}
\thanks{Email: zwzhou@ustc.edu.cn}
\affiliation{CAS Key Laboratory of Quantum Information, University of Science and Technology of China, Hefei, 230026, People’s Republic of China}
\affiliation {Synergetic Innovation Center of Quantum Information and Quantum Physics, University of Science and Technology of China, Hefei, Anhui 230026, China}
\author{Ming Gong}
\thanks{Email: gongm@ustc.edu.cn}
\affiliation{CAS Key Laboratory of Quantum Information, University of Science and Technology of China, Hefei, 230026, People’s Republic of China}
\affiliation {Synergetic Innovation Center of Quantum Information and Quantum Physics, University of Science and Technology of China, Hefei, Anhui 230026, China}
\date{\today}

\begin{abstract}
    The Haldane phase represents one of the most important symmetry protected states in modern physics. This state can be realized using spin-1 and spin-${1\over 2}$ Heisenberg models and 
bosonic particles. Here we explore the emergent Haldane phase in an alternating bond $\mathbb{Z}_3$ parafermion chain, which is different from the previous proposals from 
fundamental statistics and symmetries. We show that this emergent phase can also be characterized by a modified long-range string order, as well as four-fold 
degeneracy in the ground state energies and entanglement spectra. This phase is protected by both the charge conjugate and parity symmetry, and the edge modes are shown to satisfy parafermionic 
    statistics, in which braiding of the two edge modes yields a ${2\pi \over 3}$ phase. This model also supports rich phases, including topological ferromagnetic parafermion (FP) phase, 
trivial paramagnetic parafermion phase, classical dimer phase and gapless phase. The boundaries of the FP phase are shown to be gapless and critical with central charge $c = 4/5$. 
Even in the topological FP phase, it is also characterized by the long-range string order, thus we observe 
a drop of string order across the phase boundary between the FP phase and Haldane phase. These phenomena are quite general and this work opens a new way for finding exotic topological 
    phases in $\mathbb{Z}_k$ parafermion models.
\end{abstract}
\maketitle

Topological phases and associated phase transitions beyond Landau paradigm of phase transition have been a major topic in modern physics\cite{wen2004quantum, wen2017colloquium, witten2016fermion}.
These transitions are characterized by integer numbers in topological insulators\cite{hasan2010colloquium, qi2011topological}, or topological orders in symmetry protected topological 
phases\cite{wen2017colloquium, wen1990topological}. These phases can be used to realize exotic excitations with Abelian or non-Abelian statistics, which are building blocks for topological quantum computation. Along this line, the self-Hermitian Majorana zero modes \cite{kitaev2001unpaired} have been realized in 
experiments by several groups\cite{alicea2011non,mourik2012signatures,das2012zero,deng2012anomalous,he2017chiral,zhang2017quantized}. Moreover, the symmetry protected Haldane phase
has been proposed to be realized in spin-1 and spin-${1\over 2}$ Heisenberg models\cite{haldane1983continuum,haldane1983nonlinear,hida1992crossover,hida1992ground}, and strong interacting bosonic  
models \cite{dalla2006hidden,berg2008rise,rossini2012phase,batrouni2013competing,ejima2014spectral,lange2017anyonic}, and even fermionic models\cite{bois2015phase, capponi2016phases, barbiero2017non, nakagawa2017symmetry}. 

Here we are interested in the realization of the Haldane phase using $\mathbb{Z}_3$ parafermions\cite{fendley2012parafermionic,jermyn2014stability,zhuang2015phase,stoudenmire2015assembling,barkeshli2014synthetic,mong2014universal,vaezi2014fibonacci,alicea2016topological}, which are totally different from the above-mentioned 
models from their fundamental statistics and symmetries. We consider an alternating bond $\mathbb{Z}_3$ parafermion model, which can be realized in superconductor and $\nu={2\over 3}$ fractional quantum Hall (FQH) hybrid structure\cite{mong2014universal,vaezi2014superconducting}. This Haldane phase is characterized by nonlocal string order and four-fold degeneracy in both ground state energies and entanglement spectra.
We map out the whole phase diagram and find other exotic phases: ferromagnetic parafermion (FP) phase, paramagnetic parafermion (PP) phase, dimer phase and critical gapless phase. The boundaries for the 
FP phase are critical with central charge $c={4\over 5}$. The edge modes in the Haldane phase exhibit fractional Abelian braiding, which are protected by charge conjugate symmetry and parity symmetry.
This kind of emergent phenomena are rather general in parafermion models, thus we open a new avenue in searching of exotic symmetry protected topological phases in parafermion models.

\begin{figure}
    \centering
    \includegraphics[width=0.40\textwidth]{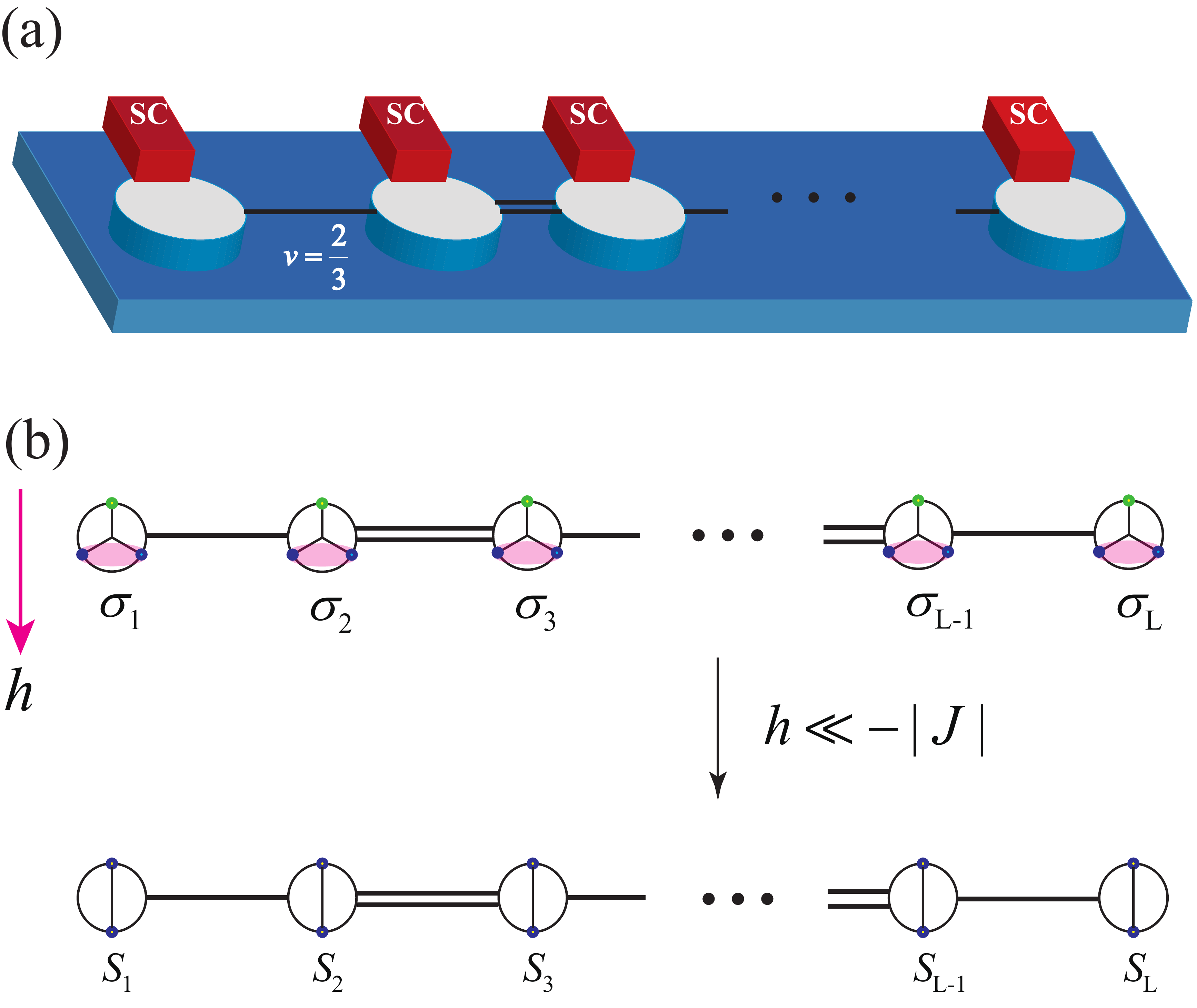}
    \caption{(a) Realization of Eq. \ref{eq-gparafermion} using superconductor and $\nu= {2\over 3}$ FQH hybrid structure. The distance between the 
    superconducting islands controls the coupling strengths between the parafermions localized at the holes, giving rise to the alternating bond model. 
    Other ideas for realizing of this model can be found in Refs. \cite{clarke2013exotic,stoudenmire2015assembling,mong2014universal,milsted2014commensurate}. 
    (b) Projection of the alternating bond $\mathbb{Z}_3$ model to the spin-${1\over 2}$ spin model in strong Zeeman field limit, $h \ll -J$.}
    \label{fig-fig1}
\end{figure}

{\it Model}. We consider the following alternating bond $\mathbb{Z}_3$ parafermion chain ($\omega^3 =1$),
\begin{eqnarray}
    H = -\omega^{2}(\sum_{j=1}^{L} J_j(\delta)\ \alpha_{2j}^{\dagger}\alpha_{2j+1} + h\ \alpha_{2j-1}^{\dagger}\alpha_{2j}) + \text{h.c.}
    \label{eq-gparafermion}
\end{eqnarray}
where $J_j(\delta) = 1 + (-1)^{j}\ \delta$ and $\alpha_j$ are parafermions satisfying $\alpha_j^3 = 1$, $\alpha_j^\dagger = \alpha_j^2$ 
and $\alpha_i \alpha_j = \alpha_j\alpha_i \omega^{\text{sgn}(i-j)}$. We propose to realize this model using the setup in Fig. \ref{fig-fig1}a in superconductor
and FQH hybrid structure\cite{alavirad2017z}, in which the alternating bonds are controlled by the distance between the superconducting islands. This model is related to the following 
$\mathbb{Z}_3$ clock model via Jordan-Wigner transformation: $\alpha_{2j -1} = \prod_{k \le j-1} \tau_k \sigma_j$ and $\alpha_{2j} = \omega \sigma_j \prod_{k \le j} \tau_k$, which yields\cite{mong2014universal,vaezi2014superconducting},
\begin{equation}
H = - \sum_{i=1}^L [1 + (-1)^{j} \delta] \sigma_i^\dagger \sigma_{i+1}  - h \sum_i \tau_i + \text{h.c.}.
    \label{eq-clock}
\end{equation} 
We see that the second term plays the similar role as the magnetic field in $\mathbb{Z}_2$ spin models, thus $h$, for convenience, is termed as Zeeman field. In 
above, the operators satisfy, $\sigma_i^3 =\tau_i^3 =1$, $\sigma_i\tau_i = \omega \sigma_i\tau_i$, $\sigma_i^\dagger = \sigma_i^2$ and $\tau_i^\dagger = \tau_i^2$; 
all operators commute between different sites\cite{fendley2012parafermionic}. 

The $\mathbb{Z}_2$ symmetry in Heisenberg spin models is absent here, which is one critical difference between the $\mathbb{Z}_2$ spin models and $\mathbb{Z}_k$ 
parafermion models. However, this symmetry, can be recovered by a projection from $\mathbb{Z}_3$ to $\mathbb{Z}_2$ model, when $h$ is negative enough, in 
which case the highest state in each site is unoccupied. In this condition, the above model may be projected to the following $\mathbb{Z}_2$ spin model to the leading term,
\begin{equation}
    H = -\sum_{i} [1 + (-1)^{j} \delta] s_i^\dagger s_{i+1},
    \label{eq-XXmodel}
\end{equation}
where $s_i = s_i^x - i s_i^y$, with $s_i^\alpha$, are Pauli matrices. This model has been investigated in literatures\cite{hida1992crossover,tzeng2016entanglement,wang2013topological}, which supports Haldane phase 
when $\delta > 0$ and trivial dimer phase when $\delta < 0$. We will show that this topological phase can be realized even with modest Zeeman field $h$. This limiting case
enables us to understand how the Haldane phase and related symmetries can emerge from the $\mathbb{Z}_3$, or even $\mathbb{Z}_k$ parafermions.

\begin{figure}
    \centering
    \includegraphics[width=0.45\textwidth]{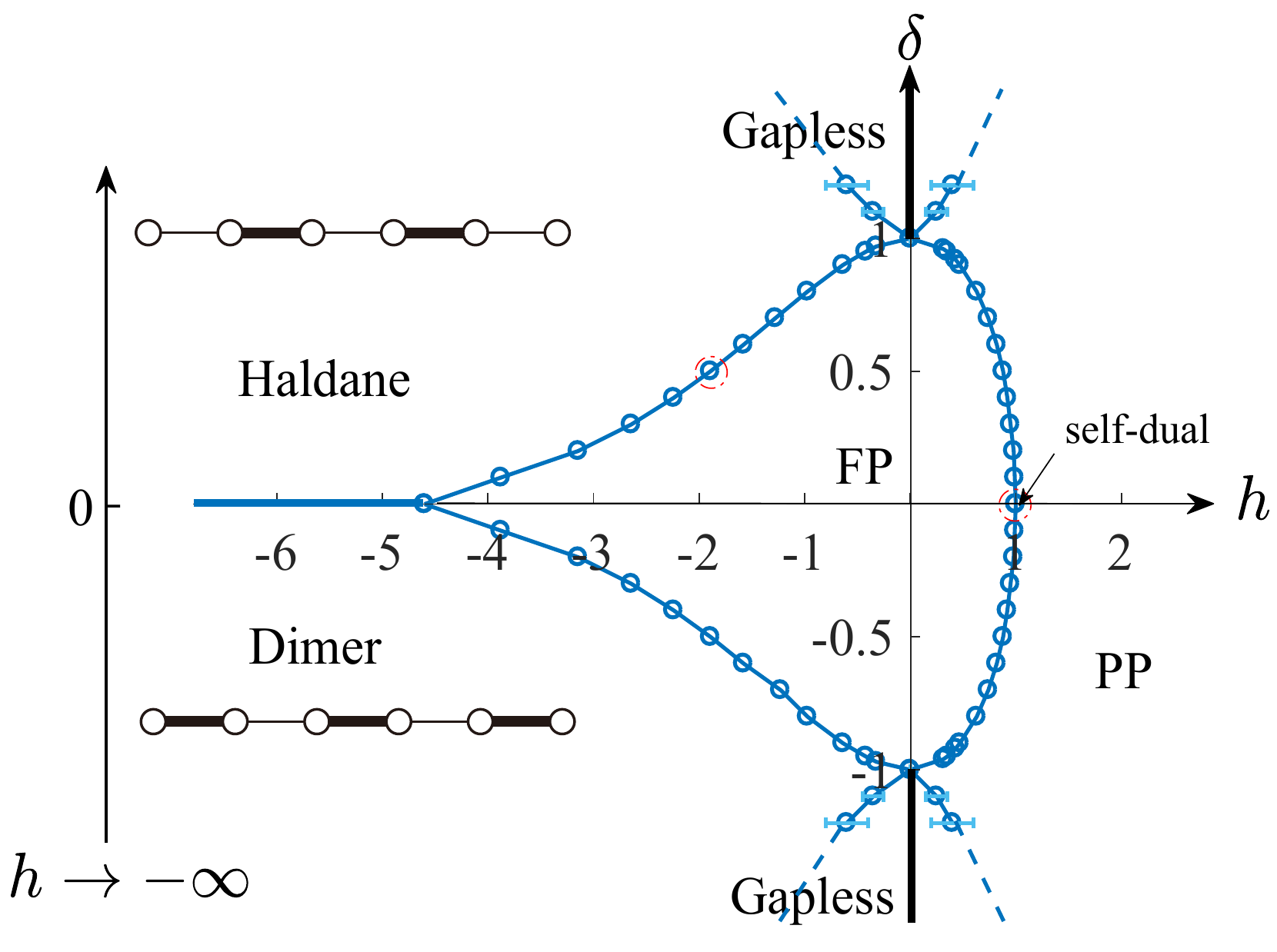}
    \caption{Phase diagram of the alternating bond parafermion model. FF and PP represent ferromagnetic and paramagnetic parafermnion phase. 
        The gapless phase with $|\delta| > 1$, $h \ne 0$ is critical with central charge $c=1$.}
    \label{fig-fig2}
\end{figure}

{\it Phase diagram}. We employ the density matrix renormalization group (DMRG) method implemented by ITENSOR project \cite{itensorcode} and exact diagonalization (ED) method to understand the phase 
diagram in Fig. \ref{fig-fig2}. Firstly, we look at $\delta = 0$, which supports several different critical points. The point at $h=J$ is self-dual\cite{fendley2012parafermionic,li2015criticality} via
$\mu_j = \prod_{k\le j} \tau_k$ and $\nu_j = \sigma_j^\dagger \sigma_{j+1}$, which separate the three-fold degenerate FP phase from 
the PP phase. This boundary was studied in literatures\cite{fendley2012parafermionic,clarke2013exotic}. When $h < 0$, we find another critical point at $h_c = -4.6$. 
When $h < h_c$, the model is equivalent to free fermion model or XX spin model, which is critical with central charge $c=1$\cite{peschel2004entanglement,ginsparg1988applied,calabrese2004entanglement,alavirad2017z,pollmann2009theory}. This point was unveiled in Ref. \cite{zhang2018topological}. On the 
other hand when $h = 0$, the corresponding clock model is purely classical, which supports critical points at  $\delta = \pm 1$. At these two points, the chain is divided
into short segments with length $\mathscr{L}=4$, giving rise to an infinite-fold degenerate model. Similar infinite-fold degeneracy can be found when $|\delta| > 1$ with
eigenvectors as $|m_1m_1m_2m_2m_3m_3\cdots\rangle$ with $m_i \ne m_{i+1}$. This picture is similar to that in the extended parafermion model\cite{zhang2018topological}. 

The line with $\delta = \pm 1$ is essential for us to understand the phases in our model. We first focus on $\delta = + 1$. In this case the coupling between the parafermions is 
decoupled into the following segments,
\begin{equation*}
    \alpha_1-\alpha_2\quad \alpha_3-\alpha_4=\alpha_5-\alpha_6 \quad \alpha_7-\alpha_8=\alpha_9-\alpha_{10}\quad \cdots
\end{equation*}
where the single and double bonds represent coupling strengths by $h$ and $J$, respectively. The two edge modes are fully decoupled from all the other sites. When $h > 0$, the ground
states for $ h \omega^2 \alpha_1^\dagger \alpha_2 + \text{h.c.} = -h (\tau + \tau^\dagger)$ is unique, corresponding to the PP phase. In contrast, when $h < 0$, the ground
states are two-fold degenerate. For this reason, it corresponds to the Haldane phase with four-fold degeneracy, taking into accounts the contribution from both ends. This 
picture is modified for the case with $\delta = -1$, which is decoupled into segments,
\begin{equation*}
    \alpha_1-\alpha_2=\alpha_3-\alpha_4 \quad \alpha_5-\alpha_6=\alpha_7-\alpha_8 \quad \alpha_9-\alpha_{10}= \cdots
\end{equation*}
We only need to compute the eigenvalues of $H_4 = -h \omega^2 \alpha_1^\dagger \alpha_2 - h \omega^2 \alpha_3^\dagger \alpha_4 - 2J\omega^2 \alpha_2^\dagger \alpha_3 + \text{h.c.}$. We always find the ground
state of $H_4$ to be unique, regardless of the sign of $h$. When $h \gg J$ it is $|0\rangle^{\otimes L}$. For $h \ll -\vert J \vert$, it is $\prod_{i} (|(2i)_2(2i+1)_1\rangle + |(2i)_1(2i+1)_2\rangle 
+ c|(2i)_0(2i+1)_0\rangle)$, where $|0\rangle$,$|1\rangle$,$|2\rangle$ are eigenvectors of $\tau$ with $\tau |i \rangle = \omega^i |i \rangle$ and $c = {1 \over 2} [\sqrt{36h^2-12h+9}+ 6h-1] \simeq 1/(3|h|) 
\ll 1$. This regime is termed as classical dimer phase, following Refs. \cite{hida1992crossover,tzeng2016entanglement,wang2013topological}, which is topological trivial.

\begin{figure}
    \centering
    \includegraphics[width=0.45\textwidth]{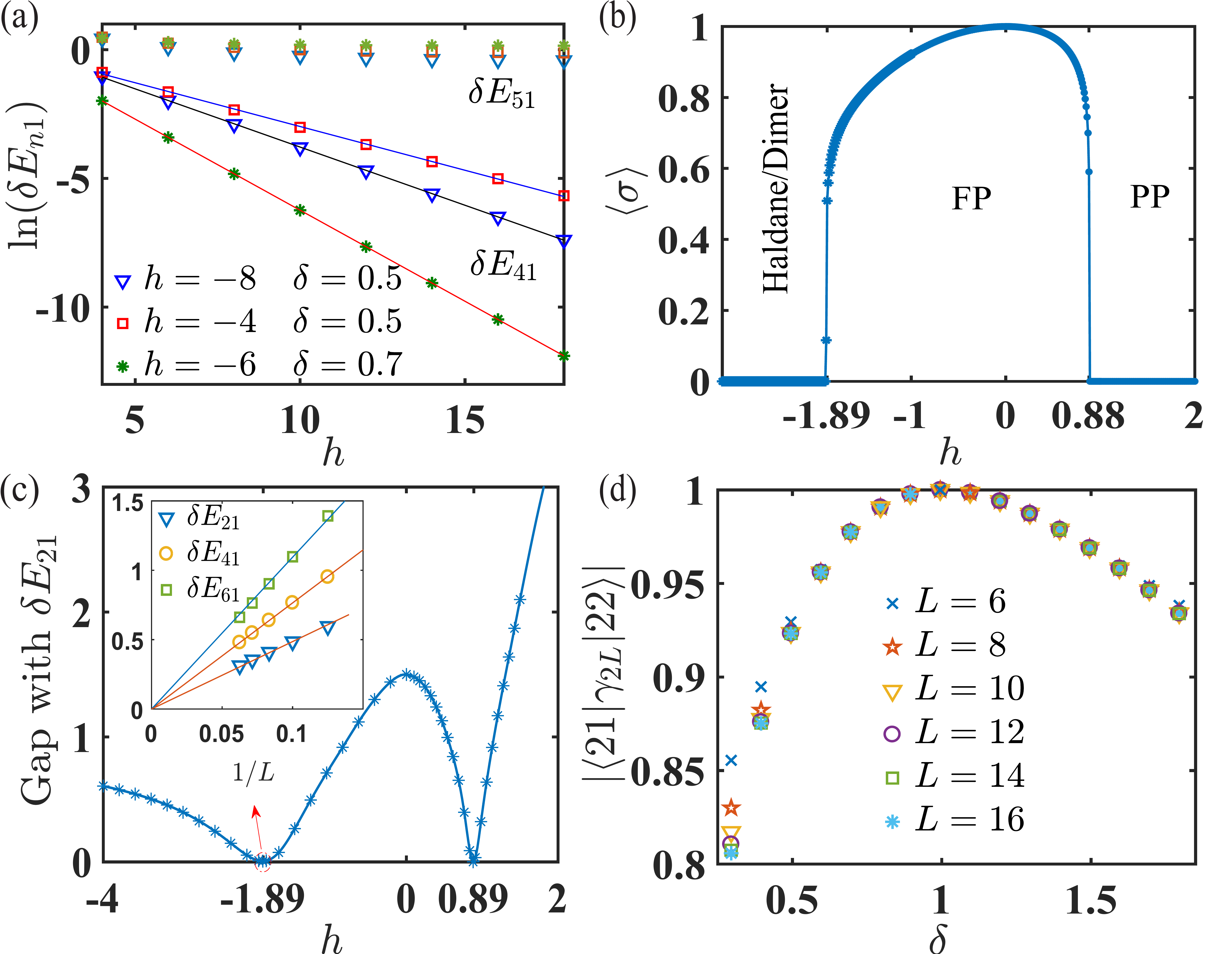}
    \caption{(a) Level spacing $\delta E_{n1} = E_n - E_1$ with open boundary condition as a function of chain length in the emergent Haldane phase. 
    (b) Ferromagnetic order $\Delta = |\langle \sigma \rangle|$ at $\delta = 0.5$ to characterize the boundaries of FP phase. (c) Energy gap $\delta E_{21}$ in the 
	infinite volume with close boundary condition. Inset show the scaling of $\delta E_{n1}$ at the critical point. (d) Overlap between edge modes and $\gamma_{2L}$ with different chain lengths for $h = -8$.}
    \label{fig-fig3}
\end{figure}

{\it Emergent Haldane phase and topological transition}. Here we mainly focus on the properties of the emergent Haldane phase. From the simple picture at $\delta = 1$, $h < 0$, we 
see that the ground states are four-fold degenerate. We consider a more general case in Fig. \ref{fig-fig3}a for a finite system with open boundary condition based on ED method. The 
splitting between the fourth level and ground state, $\delta E_{41} = E_4 - E_1$, decreases exponentially to zero with the increasing of length $L$, indicating of the exact four-fold degeneracy in infinite volume. This phase is protected by a finite energy gap between the fifth and ground state levels, i.e., $\lim_{L\rightarrow \infty}\delta E_{51}(L)$ is finite. With close boundary condition \cite{alexandradinata2016parafermionic,motruk2013topological}, the edge modes are paired, leading to a unique ground state. In this case we perform the same analysis in Fig. \ref{fig-fig3}c, and show that the gap $\delta E_{21} = E_2 - E_1$ vanishes only at the phase boundaries. We show in the inset that at the phase boundary, $\delta E_{n1} \propto 1/L$ for all $n> 1$ with finite $L$, due to criticality. 

We determine the phase boundaries for the FP phase using the ferromagnetic order $\Delta = |\langle \sigma \rangle|$, which is realized in numerical simulation without
$\mathbb{Z}_3$ symmetry restriction; see details in Ref. \cite{zhang2018topological}. This method enables us to precisely determine the phase boundaries due to the sharp transitions
from finite in FP phase to zero in all other phases, including that in Haldane phase and dimer phase. Notice that the case for $\delta$ and $-\delta$ give the same value for $\Delta$,  
as expected, since these two cases can be made the same upon one lattice translation.

We now discuss the particular properties of the edge modes, which show some distinct features as compared with those
from $\mathbb{Z}_2$ spin models. Let us again consider $\delta = +1$ and $h < 0$. In this case the left two parafermions, $\alpha_1$, $\alpha_2$, and right two parafermions,
$\alpha_{2L-1}$, $\alpha_{2L}$, are decoupled from the bulk $H(\alpha_3, \alpha_4, \cdots, \alpha_{2L-3}, \alpha_{2L-2})$. Taken the left two parafermions as an example, we may 
 represent $\alpha_1 = \sigma$ and $\alpha_2 = \omega \sigma \tau$, then the Hamiltonian can be written as $H = -h (\tau + \tau^\dagger) = -h\text{diag}(2, -1, -1)$. Let us define left(right) projector $P_l(P_r) = 
|1\rangle\langle 1| + |2\rangle \langle 2| = \text{diag}(0, 1, 1)$, which project any wave function to the lowest two eigenvectors of $H$, then we can approximate the two edge modes as 
$\alpha_i\rightarrow \gamma_i = P_l \alpha_i P_l$. We find that $\gamma_{1} = \gamma_2$, which are zero modes of the original Hamiltonian, {\it i.e.}, $[\gamma_1, H] = 0$. 
The similar results can be found for the right zero modes for $\gamma_{2L} = P_r\alpha_{2L} P_r $. These two modes satisfy fermionic relation at the same site, 
$\gamma_1^2= \gamma_{2L}^2 = 0$, and parafermionic commute relation between the two ends, $\gamma_1 \gamma_{2L}=  \omega \gamma_{2L} \gamma_1$, which mark the important difference between our model and the previous fermionic edge modes \cite{su1979solitons}. If we denote this state as $|11\rangle$, then we can construct three other zero modes as 
$|12\rangle = \omega\gamma_{2L}^\dagger |1,1\rangle$, $|21\rangle = \gamma_{1}^\dagger |11\rangle$, $|22\rangle = \omega^2\gamma_{2L}^\dagger |21\rangle$. 

The zero modes are exact only in the limiting case of $\delta = 1$, $h<0$. However, we expect these edge modes to be localized over a wide region, which extend to a 
distance of the order of one correlation length from the open ends\cite{motruk2013topological}. In this case, the exact edge operators at the two ends have a 
finite overlap with $\gamma_1$ and $\gamma_{2L}$. To this end  we diagonalize the Hamiltonian with open boundary condition at $h=-8$, and calculate the matrix element 
of $\gamma_{2L}$ between $|22 \rangle$ and $|21 \rangle$. This overlap depends only on the correlation length when $L$ is large enough. For the result in Fig. \ref{fig-fig3}d, 
the correlation length $\xi \sim 1$ when $\delta \rightarrow 1$, thus the overlap is saturated even in a short chain. 

\begin{figure}
    \centering
    \includegraphics[width=0.45\textwidth]{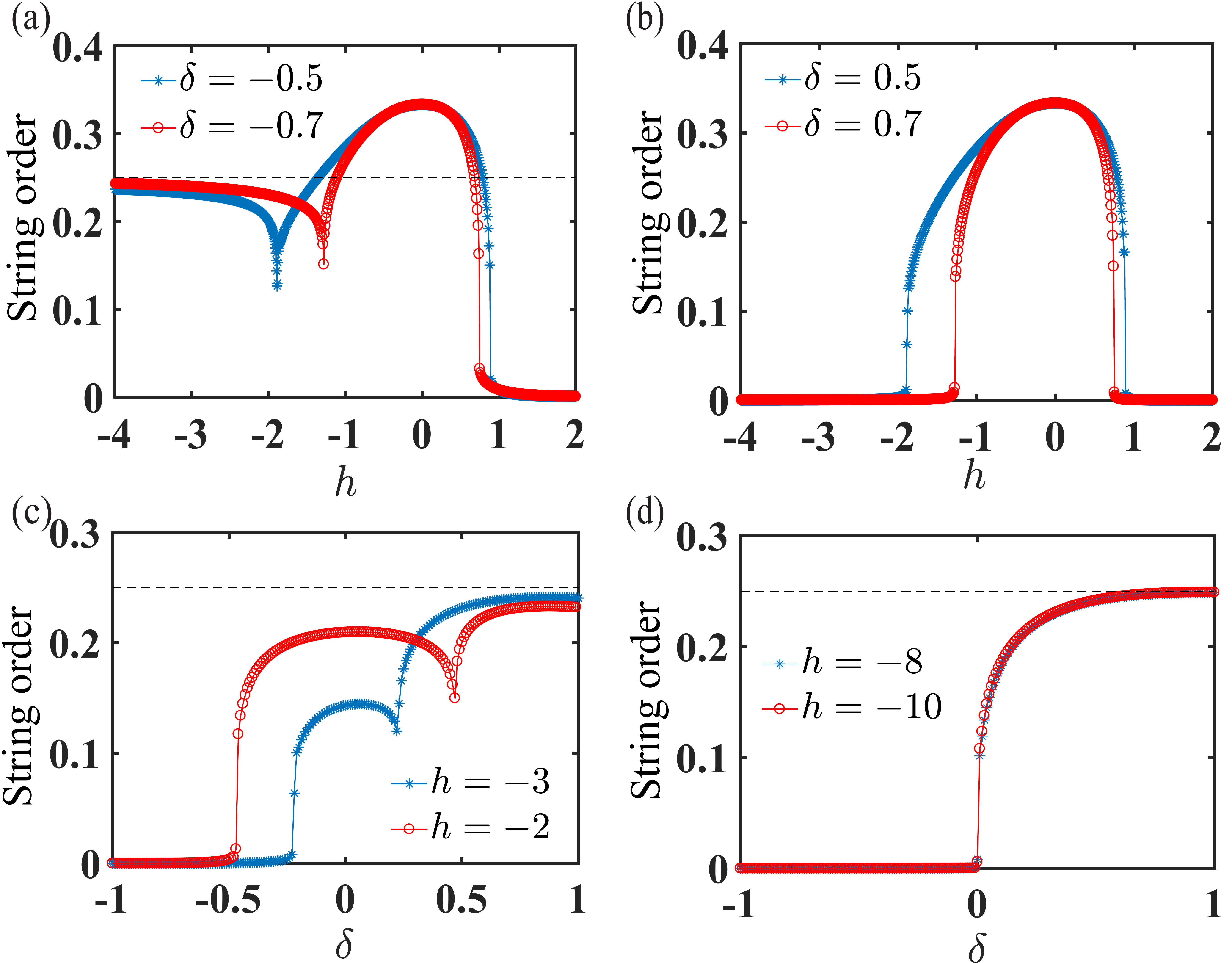}
    \caption{Long-range string order as a function of $\delta$ (a-b) and $h$ (c-d). Results are obtained using infinite-chain DMRG method with bond dimension $m = 200$. 
    The horizontal dashed lines mark the limit of $\mathbb{Z}_2$ spin model with $O_\text{s} = O_\text{s}^{\text{xx}} = {1\over 4}$.}
    \label{fig-fig4}
\end{figure}

\begin{figure}
    \centering
    \includegraphics[width=0.45\textwidth]{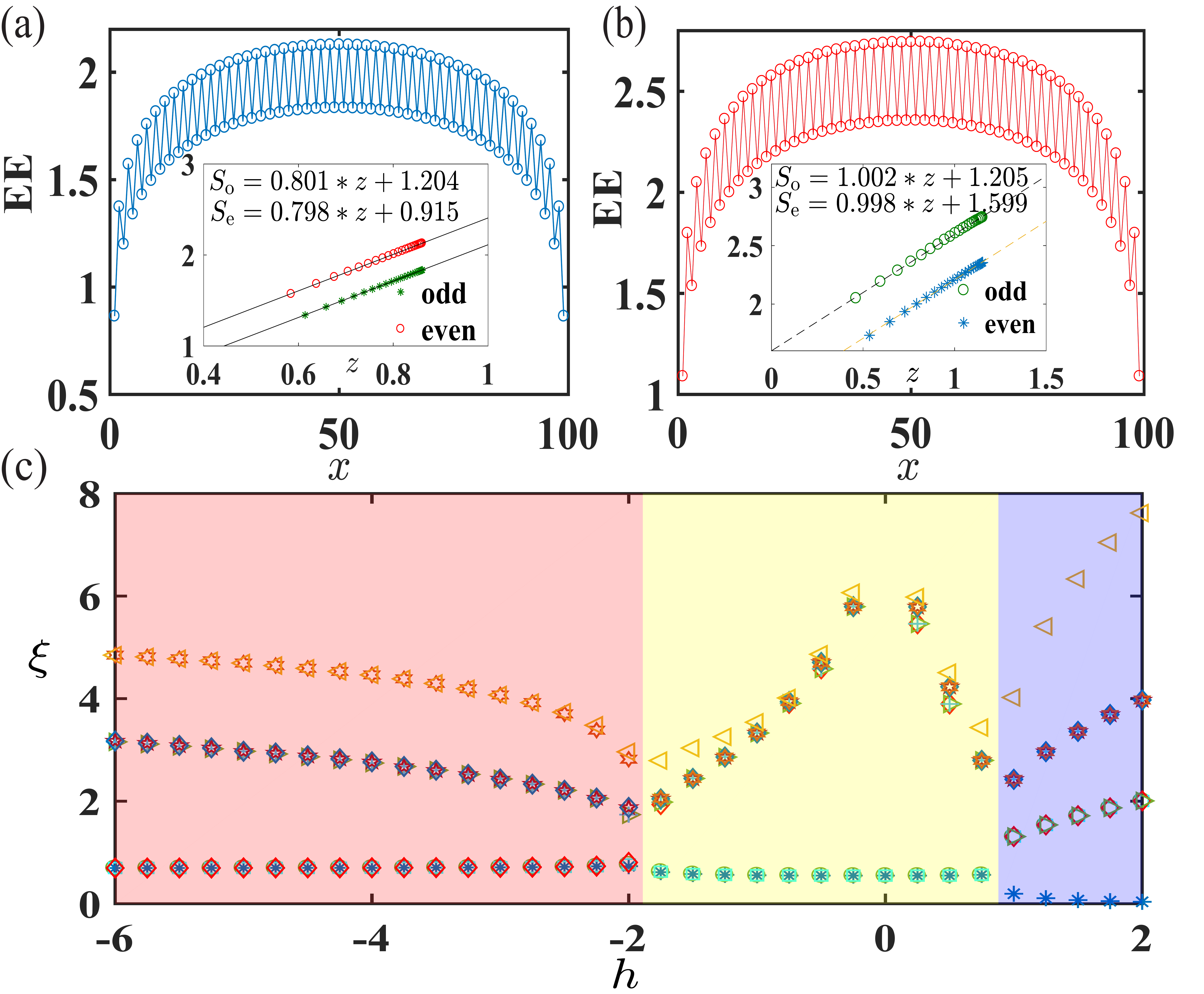}
    \caption{Von Neumann entanglement entropy in the phase boundary of FP phase with $h=-1.89$, $\delta=0.5$ (a) and C phase with $h=-0.22$, $\delta=1.1$ (b). Results are obtained
    using periodic boundary condition with $m = 2000$. Inset show corresponding fitted central charge for the even sites and odd sites, respectively. In horizontal axis, we set $z=\frac{1}{3} \text{ln} ({L \over \pi} \sin(\pi {x \over L}))$. (c) Entanglement spectra in Haldane phase, FP phase and PP phase. Results are obtained for $L=80$, $\delta=0.5$, $m=2000$
    with periodic boundary condition.}
    \label{fig-fig5}
\end{figure}

{\it Long-range string order}. We define the string order as
\begin{eqnarray}
    O_{{\rm s}} = \lim_{\vert i-j\vert \rightarrow \infty} \langle \sigma_{2i} U_{2i+1} U_{2i+2} \cdots U_{2j-2} \sigma_{2j-1}^\dagger \rangle. 
    \label{eq-str}
\end{eqnarray}
which is generalized from the string order in previous  Haldane phases\cite{den1989preroughening,kennedy1992hidden1,kennedy1992hidden2,perez2008string,pollmann2012detection}. The center operator $U_i$ in Eq.\ \ref{eq-str} is the generator of the charge conjugate symmetry, satisfying $U_i\tau_i U_i^\dagger =\tau_i^\dagger$ and $U_i\sigma_i U_i^\dagger =\sigma_i^\dagger$. This string order is not unique, nevertheless, the one in Eq.\ \ref{eq-str} is enough to distinct the topological states from the trivial phases. One should be noticed that when transform to the parafermion picture, the above string order can be written as $O_{{\rm s}} = \lim_{\vert i-j\vert \rightarrow \infty} \langle \psi_{2i}(U_{2i+1}\tau_{2i+1}) \cdots (U_{2j-2}\tau_{2j-2})\chi_{2j-1}^{\dagger}\rangle$ ($\chi_k$, $\psi_k$ are parafermions $\alpha_{2k-1}$,$\alpha_{2k}$ , and $\tau_k$ transforms as $\chi_k^{\dagger}\psi_k$), which is also made by a nonlocal string. Thus after the Jordan-Wigner transformation, the Haldane phase will turn to a new phase, which possesses all the necessary properties for defining a symmetry protected topological Haldane phase. For this reason, even in parafermion picture, this phase is also called as Haldane phase. In Fig.\ \ref{fig-fig4}, we compute the string order using infinite chain DMRG method\cite{mcculloch2008infinite}, sweeping along both 
horizontal and vertical directions on the phase diagram. We find that this string order in the topological phases is nonzero, and drops to zero in all trivial phases. From Fig.\ \ref{fig-fig4}a-b, we see that when sweeping from the Haldane phase to FP phase and PP phase, the string order undergos a sudden drop at the phase boundary between FP and Haldane phase, which was originally determined by ferromagnetic order $\Delta$. In the PP phase, it drops to zero. This is different from the case with $\delta < 0$, in which the string order is zero in both the dimer and PP phases. In Fig.\ \ref{fig-fig4}c-d, we plot the string order as a function of $\delta$, which also exhibits similar features. Especially, when across the boundary between dimer and Haldane phase with $h < -4.6$, the string order is shown to be independent of $h$. 
Noticed that in the strong Zeeman field limit, the string order is reduced to the following one by projecting to the lowest two eigenvectors in each sites, 
$O_{{\rm s}}^{{\rm xx}}  = -\lim_{\vert i-j\vert \rightarrow \infty} \langle s_{2i} s_{2i+1}^x s_{2i+2}^x \cdots s_{2j-2}^x s_{2j-1}^\dagger \rangle = 1/4$ \cite{hida1992ground}, where $s_i^\alpha$ are 
Pauli matrices; see Eq.\ \ref{eq-XXmodel}. 

{\it  Central charge and entanglement spectra}. We further characterize the boundaries and the gapless phase by entanglement entropy (EE).
In a finite chain with periodic boundary condition, it can be written as\cite{calabrese2009entanglement,calabrese2004entanglement,vidal2003entanglement,li2015criticality,ginsparg1988applied}
\begin{equation}
        S(x) \sim {c \over 3} \ln ({L \over \pi} \sin {\pi x \over L}), \quad
        \label{eq-EE}
\end{equation}
where $c$ defines the central charge. For the phase boundaries determined by the ferromagnetic order, we find that $c = {4\over 5}$. One should be noticed that due to the 
alternating bond strengths between the neighboring sites, the EE also exhibits oscillating behaviors, thus Eq.\ \ref{eq-EE} should be fitted for the odd and even sites, respectively. 
With this technique, we find $c = 1$ in gapless phase, after breaking of infinite-fold degeneracy by the Zeeman field. 

We provide more insight into the physics in these two topological phases from the entanglement spectra\cite{li2008entanglement,pollmann2010entanglement,motruk2013topological,fidkowski2010entanglement,fidkowski2011topological,turner2011topological,hsieh2014bulk,wang2018decoding}, which is defined as $\xi_i = -\text{ln}(\rho_i) $. Here, $\rho_i$ is the eigenvalue of the reduced density matrix $\hat{\rho}_A = \text{Tr}_B \vert \psi \rangle \langle \psi \vert $, where $\vert \psi \rangle $ is the ground-state wave function,
and $A,B$ are two partitions of the parafermion chain. As was unveiled in Ref. \cite{motruk2013topological}, the spectra $\xi_i$ is three-fold degenerate in FP phase. 
However, in Haldane phase, it is characterized by four-fold degeneracy\cite{turner2011topological}. Our results are presented in Fig.\ \ref{fig-fig5}c, in which we 
used a closed parafermion chain \cite{alexandradinata2016parafermionic,motruk2013topological} with chain length $L=80$ and the chain is cutted from the center for two partitions. We find that for the lowest eight eigenvalues, they are strictly four-fold degenerate in the Haldane phase. Nevertheless, for the higher spectra, their values are vanishing small, but no longer four-fold degenerate. The whole spectra are exactly four-fold degenerate in the limit when Zeeman field is negative enough. In FP phase, all the spectra are three-fold degenerate, in agreement with the previous observation \cite{motruk2013topological}. In the PP phase, which is trivial, all spectra are not degenerate anymore. 

{\it Topological protection}. We find that the Haldane phase is protected by both charge conjugate symmetry $C = \prod_i U_i \mathcal{K}$, where $U_i$ was defined in Eq.\ \ref{eq-str} and  $C^2 = +1$, and $\mathbb{Z}_3$ parity $P = \prod_i \tau_i$, with $P^3 = 1$. The four-fold degeneracy will break down only for terms violating these two symmetries. Let us define two left(right) edge states, where $C|1\rangle = \omega |1\rangle$ and $C|2\rangle = \omega^2 |2\rangle$. By projecting to these two edge modes, we find the projective representation: $\mathcal{P}_L = P_{L} P P_{L} = \begin{pmatrix}\omega & 0 \\ 0  & \omega^2\end{pmatrix}$ and $\mathcal{C}_L = P_{L} C P_{L} = \begin{pmatrix} 0 & 1 \\ 1  & 0\end{pmatrix} \mathcal{K}$, which satisfy $[H, \mathcal{P}_L \mathcal{C}_L] = 0$ in the ground state subspace. We find that for any eigenvector $|\psi\rangle$ in the subspace, $\mathcal{P}_L \mathcal{C}_L|\psi\rangle$ is impossible to be identical to $|\psi\rangle$, since $(\mathcal{P}_L \mathcal{C}_L)^2 = \mathcal{P}_L^* \ne 1$. For this reason, in each open end, the ground states should be two-fold degenerate, giving rise to the four edge modes $|ij\rangle$, with parity $\omega^{i+j}$. This idea also applies well to other parafermion models.

{\it Conclusion}. To conclude, we investigate the emergent Haldane phase in the $\mathbb{Z}_3$ alternating bond parafermion chain, which exhibits four-fold degeneracy in both ground states and entanglement spectra, and is characterized by nonzero long-range string order, generalized from the $\mathbb{Z}_2$ spin models. Strikingly, we find 
that the FP phase is also characterized by nonzero string order and three-fold degeneracy in both ground states and entanglement spectra, which separate from the Haldane 
phase by a critical phase with central charge $c={4\over 5}$. With our construction, the similar phases can also be found in other $\mathbb{Z}_k$ parafermion chains, in
regarding of the rather generality of emergent phenomena in these models. We expect the results in this work open a new way for finding of exotic topological phases in $\mathbb{Z}_k$ parafermion models.

\begin{acknowledgments}
        {\it Acknowledgement}. M.G. is supported by the National Youth Thousand Talents Program (No. KJ2030000001), the USTC start-up funding (No. KY2030000053) and the NSFC (No. GG2470000101). Z. Z. and G. G. are supported by National Key Research and Development Program (No. 2016YFA0301700), National Natural Science Foundation of China(No. 11574294), and the ''Strategic Priority Research Program (B)'' of the Chinese Academy of Sciences (No. XDB01030200).
\end{acknowledgments}

\bibliography{ref}

\end{document}